\documentclass[12pt,subeqn,epsbox]{article}
\usepackage{graphicx}
\usepackage{amsmath}

\def\beq{\begin{equation}}
\def\eeq{\end{equation}} 
\def\beqa{\begin{eqnarray}} 
\def\eeqa{\end{eqnarray}} 
\def\bfig{\begin{figure}\vspace{5mm}}

\def\efig{\end{figure}}

\def\bqu{\begin{quote}}
\def\equ{\end{quote}}
\def\bitem{\begin{itemize}}
\def\eitem{\end{itemize}}
\def\ben{\begin{enumerate}}
\def\een{\end{enumerate}}

\def\order{{\cal O}}

\def\<{\langle}
\def\>{\rangle}
\def\sbeqa{\begin{subequations}}
\def\seeqa{\end{subequations}} 
\def\non{\nonumber}
\def\xis{\xi^{{\rm s}}}
\def\bxis{\bar{\xi}^{{\rm s}}}

\title{Method of Collective Degrees of Freedom in Spin Coherent State Path Integral\footnote{
"XXIII International Colloquium on Group Theoretical Methods in Physics: Dubna 2000, July 31- August 5, 2000, Dubna, Russia" to be published in 
Phys. Atom. Nucl., {\bf 12}, (2001) as  
proceedings of the topical section of "Symposium on group theory and 
path integrals" 
}}
\author{Junya Shibata$^a$ and Shin Takagi$^b$\\
$^a$Department of Physics, Tohoku University, 
Sendai 980-8578, Japan,\footnote{present address:
Department of Physics, Osaka University, 
Toyonaka, Osaka 560-0043, Japan, shibata@acty.phys.sci.osaka-u.ac.jp}\\
$^b$Fuji Tokoha University, Fuji 417-0801, Japan.}
\date{}
\begin{document}
\maketitle
\begin{abstract}
We present a detailed field theoretic description
of those collective degrees of freedom (CDF)
which are relevant to study macroscopic quantum dynamics
of a quasi-one-dimensional ferromagnetic domain wall.
We apply spin coherent state path integral (SCSPI)
in the proper discrete time formalism
(a) to extract the relevant CDF's,
namely, the center position and the chirality of the domain wall,
which originate from the translation and the rotation invariances
of the system in question,
and (b) to derive effective action for the CDF's
by elimination of environmental zero-modes with the help of the {\it
Faddeev-Popov technique}.
The resulting effective action turns out to be such
that both the center position and the chirality can be formally described
by boson coherent state path
integral. However, this is only formal; there is a subtle departure from
the latter.
\end{abstract}

\section{Introduction}
Recent so-called nanostrucure technology enables us 
to study low dimensional magnetism 
in mesoscopic magnets from the quantum mechanical point of view 
[1-3]. 
Among others, a magnetic domain wall has attracted 
much attention both theoretically and experimentally, 
because it is expected to exhibit macroscopic quantum phenomena 
[4-11]. 
As a theoretical technique to evaluate the quantum dynamics of the 
domain wall, 
the spin coherent state path integral in the continuous-time 
formalism [12] is frequently used.
However, as noted by some workers [13,14], 
it has some fundamental difficulties, which have been recently 
discussed in detail [15]. 
Furthermore, 
it is liable to lead to confusion concerning the 
interpretation of the collective degrees of freedom 
as has been pointed out in Ref.[16]. 
Hence, as yet there is no microscopic theory of the 
quantum dynamics of the domain wall. 
In order to pave the way for such a theory, this paper 
presents a field theoretic description 
of collective degrees of freedom by use of 
the spin coherent state path integral in the proper discrete-time 
formalism.
 
\section{Model}

We consider a ferromagnet consisting of a spin $\mbox{\boldmath$S$}$ of magnitude ${\it S}$ at each site in a quasi-one-dimensional cubic crystal 
(a linear chain) 
of lattice constant $a$. 
The magnet is assumed to have an easy axis in the $z$ directions. 
Accordingly, we adopt the Hamiltonian 
\beqa
\hat{H} = -\tilde{J}\sum_{<i,j>}^{N_{L}}\hat{\mbox{\boldmath$S$}}_{i}
\cdot\hat{\mbox{\boldmath$S$}}_{j} 
          -\frac{K}{2}\sum_{j}^{N_{L}}\hat{S}^2_{j,z},
\label{hami1}
\eeqa
where the index $i$ or $j$ represents a lattice point, $<i,j>$ denotes 
a nearest-neighbor pair, $N_{L}$ is the total number of lattice points, 
and $\tilde{J}$ is the exchange coupling constant, 
and $K$ is the longitudinal anisotropy constants; 
$\tilde{J}$, $K$ are all positive.

Since we are interested in those transition amplitudes 
which are appropriate to describe quantum mechanical motion of a domain wall, 
we introduce a {\it spin-coherent state} [17] at each site, 
which is suited for a vector picture of spin. 
We denote a state of the system as 
\beqa
|\xi\> \equiv |\xi_{1},\xi_{2},\cdots,\xi_{N_{L}}\>
        := \bigotimes_{j}^{N_{L}}|\xi_{j}\>,
\label{total-xi}
\eeqa
where $|\xi_{j}\>$ is 
a spin coherent state at the site $j$.
The transition amplitude between the initial state $|\xi_{{\rm I}}\>$ and the final 
state $|\xi_{{\rm F}}\>$ can be expressed as a 
{\it spin-coherent-state path integral} in the real {\it discrete-time} 
formalism by the standard procedure of the repeated use of the resolution 
of unity (see, e.g., Ref.15 on which the present notation is based):
\beqa
\< \xi_{{\rm F}}|e^{-i\hat{H}T/\hbar}|\xi_{{\rm I}} \>
= \lim_{N\to\infty}\int\prod_{n=1}^{N-1}\prod_{j}^{N_{L}}
d\mu(\xi_{j}(n),\xi^{*}_{j}(n))
\exp\left(\frac{i}{\hbar}{\cal S}[\xi^{*},\xi]\right),
\label{ta1}
\eeqa
where $N \equiv T/\epsilon$, $\epsilon$ is an infinitesimal time interval, 
$n$ represents discrete time, and the integration measure is 
\beqa
d\mu(\xi,\xi^{*}) 
:= \frac{2S+1}{(1+|\xi|^2)^2}\frac{d\xi d\xi^{*}}{2\pi i },
\qquad\frac{d\xi d\xi^{*}}{2\pi i }\equiv \frac{d\Re{\xi}d\Im{\xi}}{\pi}.
\label{measure1}
\eeqa
The action ${\cal S}[\xi^{*},\xi]$ consists of two parts, 
${\cal S}^{{\rm c}}[\xi^{*},\xi]$ and ${\cal S}^{{\rm d}}[\xi^{*},\xi]$, 
which are to be called the {\it canonical term} and the {\it dynamical term}, respectively. We shall be interested in 
those spin configurations whose scale of spatial variation 
is much larger than the lattice constant $a$. 
Accordingly, 
we take the spatial continuum limit in the action:
\sbeqa
\label{actions}
\beqa
{\cal S}[\xi^{*},\xi] &:=& {\cal S}^{{\rm c}}[\xi^{*},\xi]
+ {\cal S}^{{\rm d}}[\xi^{*},\xi],\\
\frac{i}{\hbar}{\cal S}^{{\rm c}}[\xi^{*},\xi]
&=& S\sum_{n=1}^{N}\int_{-L/2}^{L/2}\frac{dx}{a}
\ln\frac{(1+\xi^{*}(x,n)\xi(x,n-1))^2}{(1+|\xi(x,n)|^2)(1+|\xi(x,n-1)|^2)},
\label{sconti-canonical}\\
\frac{i}{\hbar}{\cal S}^{{\rm d}}[\xi^{*},\xi]
&=&-\frac{i}{\hbar}\sum_{n=1}^{N}\epsilon \int_{-L/2}^{L/2}\frac{dx}{a}
{\cal H}(\xi^{*}(x,n),\xi(x,n-1)),
\label{sconti-dynamical}\\
{\cal H}(\xi^{*}(x),\eta(x)) 
&:=&\frac{S}{(1+\xi^{*}(x)\eta(x))^{2}}
\Bigg[ 2JS \partial_{x} \xi^{*}(x)\partial_{x}\eta(x)\non\\
&&-\frac{K}{2}\left\{\left(S-\frac{1}{2}\right)(1-\xi^{*}(x)\eta(x))^{2}
+\frac{1}{2}\right\}\Bigg],\label{sconti-hami}
\eeqa
\seeqa
where $L$ is the length of the linear chain and 
$J \equiv \tilde{J}a^{2}$. 

\section{Method of Collective Degrees of Freedom}
\subsection{Kink configuration}

We begin by finding a domain wall configuration. 
It is determined by one of the static solutions $\{ \xis(x),\bxis(x)\}$ 
of the action ${\cal S}[\xi^{*},\xi]$.
They satisfy the following equations:
\sbeqa
\label{stat-eq}
\beqa
&&\lambda^{2}\left\{\partial_{x}^{2}\xis(x)-\frac{2\bxis(x)(\partial_{x}\xis(x))^2}{1+\bxis(x)\xis(x)}\right\}-\frac{1-\bxis(x)\xis(x)}{1+\bxis(x)\xis(x)}\xis(x) =0,
\label{stat-eq1}\\
&&\lambda^{2}\left\{\partial_{x}^{2}\bxis(x)-\frac{2\xis(x)(\partial_{x}\bxis(x))^2}{1+\bxis(x)\xis(x)}\right\}-\frac{1-\bxis(x)\xis(x)}{1+\bxis(x)\xis(x)}\bxis(x) =0,
\label{stat-eq2}
\eeqa
\seeqa
where $\lambda^2 \equiv JS/K(S-1/2)$. 
An obvious solution is the "vacuum" solution representing 
the uniform configuration in which the spins are 
either all parallel or all anti-parallel to the $z$ direction. 
The other solution is the "kink" solution representing
a domain-wall configuration in which the spins at $x\sim +\infty$ 
are parallel to the $z$ direction, the spins at $x\sim -\infty$ 
are anti-parallel to the $z$ direction, and there is a transition region 
(i.e., a domain wall) of width $\lambda$; 
\beqa
\xis(x) = \exp\left(-\frac{x-Q}{\lambda} + i\phi_{0}\right),
\qquad
\bxis(x) = \exp\left(-\frac{x-Q}{\lambda} - i\phi_{0}\right),
\label{stat-sol}
\eeqa
where $Q$ and $\phi_{0}$ are arbitrary real constants. 
$Q$ is the center position of the domain wall, 
and $\phi_{0}$ is a quantitative measure of the {\it chirality} 
of the domain wall with respect to the $x$ axis (Fig. 1); 
the wall is maximally right-handed if $\phi_{0} = \pi/2$ 
and maximally left-handed if $\phi_{0}=-\pi/2$, 
while it has no chirality if $\phi_{0}=0$. 
The range of $\phi_{0}$ is chosen as $-\pi \le \phi_{0} \le \pi$, 
with $\phi_{0}=\pi$ and $\phi_{0}=-\pi$ representing the same situation. 
$\{\xis(-x), \bxis(-x)\}$ is also a solution representing 
a domain-wall configuration. 
However, this as well as the vacuum solution 
belongs to a sector different from that of (\ref{stat-sol}). 
Since a transition 
between different sectors is forbidden [18], 
it is sufficient to consider only the sector 
(\ref{stat-sol}) for the purpose of studying
 the dynamics of a domain wall. 

\begin{figure}[htbp]
\begin{center}
	\includegraphics[scale=0.6]{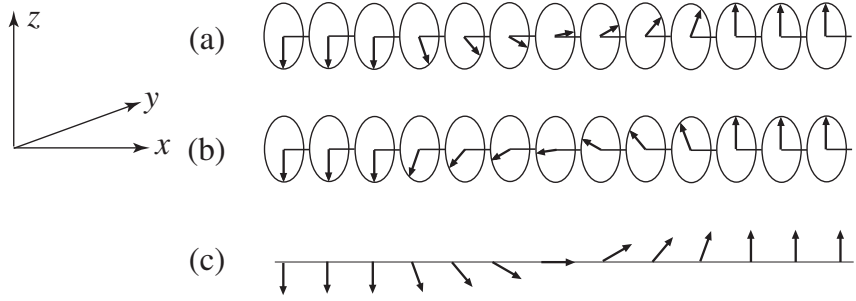}
\end{center}	
	\caption{Domain walls with three chiralities (quoted from Ref.10); (a) right-handed wall $(\phi_{0}=\pi/2)$, (b)left-handed wall $(\phi_{0}=-\pi/2)$, and (c) wall with no chirality $(\phi_{0}=0)$. Circles in (a) and (b) drawn to guide the eye lie in the $yz$ plane, while the spins lie in the $zx$ plane in (c). The quasi-one-dimensional direction of the crystal is here aligned with the spin hard axis for ease of visualization. A different alignment, which may be the case for a real magnet, does not affect the content of the text; for instance, one could rotate all the spins by $\pi/2$ around the $y$ axis if the dominant anisotropy originates from the demagnetizing field.}
	\label{Fig1}
\end{figure}

\subsection{Collective degrees of freedom and environment}

Study of the domain-wall dynamics 
is facilitated by introducing relevant collective degrees of freedom. 
We note two kinds of invariance possessed by Eq.(\ref{stat-eq}). 
One is the translation invariance in the $x$ direction.  
The other is the rotation invariance around the $z$ axis.  
These invariances are embodied by the arbitrariness in the choice of $Q$ 
and $\phi_{0}$, respectively, in (\ref{stat-sol}). 
Hence we elevate them to dynamical variables $Q(n)$  
and $\phi_{0}(n) $ [8,10,18].
To deal with these two dynamical variables (collective degrees of freedom), 
it is convenient to define 
\beqa
&&z(n) :=  q(n) + i\phi_{0}(n), \qquad 
z^{*}(n) := q(n) - i\phi_{0}(n),
\label{z-qp}\\
&&q(n)\equiv Q(n)/\lambda,\qquad n = 1,2,...,N-1. \nonumber
\eeqa
By use of these variables, original integration variables 
$\xi(x,n)$ and $\xi^{*}(x,n)$ may be decomposed 
into the domain-wall configuration and the deviation from it:
\sbeqa
\label{xi-c-e}
\beqa
\xi(x,n) &=& \xis(x;z(n)) + \tilde{\eta}(x,n;\{z\}^{n}_{n}),\\
\label{xi-c-e1}
\xi^{*}(x,n) &=& \bxis(x;z^{*}(n)) + \tilde{\eta}^{*}(x,n;\{z^{*}\}^{n}_{n}),
\label{xi-c-e2}
\eeqa
\seeqa
where
\beqa
\xis(x;z(n)) := \exp\left(-x/\lambda + z(n)\right) ,\qquad
\bxis(x;z^{*}(n)) := \exp\left(-x/\lambda + z^{*}(n)\right),
\label{stat-sol2}
\eeqa
and we use the notation;$\{z\}^{n}_{m}=\{z^{*}\}^{m}_{n}
:=(z^{*}(n),z(m))$. 
At both ends of the discrete time ($n=0$ or $n=N$), we define 
\beqa
&&z(0) \equiv z_{{\rm I}} := q_{{\rm I}}+i\phi_{{\rm I}},\qquad 
z(N) \equiv z_{{\rm F}} := q_{{\rm F}}+i\phi_{{\rm F}},
\label{z0N}\\
&&\eta(x,0;\{z\}^{0}_{0})=\eta^{*}(x,N;\{z\}^{N}_{N})=0,
\label{eta0N}
\eeqa
where $z_{{\rm I}}$ represents the center position $q_{{\rm I}}$ 
and the chirality $\phi_{{\rm I}}$ of the domain wall in the initial state, 
and $z_{{\rm F}}$ those in the final state.

The variable $\tilde{\eta}$, which are to be called environment
around the domain wall, 
could be expanded by a set of some mode function. 
The expansion is expected to contain the zero-modes, 
which originate from the translation and the rotation invariances 
and should be eventually eliminated 
in order to avoid overcounting the degrees of freedom. 
For this reason, we expand 
the environment with respect to a set of mode functions $\{\psi_{k}\}$ 
as 
\sbeqa
\label{env}
\beqa
\tilde{\eta}(x,n;\{z\}^{n}_{n}) &=& \eta_{0}(n)\psi_{0}(x;z(n))
+ \eta(x,n;\{z\}^{n}_{n}),\label{env1}\\
\tilde{\eta}^{*}(x,n;\{z^{*}\}^{n}_{n}) 
&=& \eta^{*}_{0}(n)\psi^{*}_{0}(x;z^{*}(n))
+ \eta^{*}(x,n;\{z^{*}\}^{n}_{n}),\label{env2}\\
\eta(x,n;\{z\}^{n}_{n}) &=& {\sum_{k}\/}' \eta_{k}(n)\psi_{k}(x;\{z\}^{n}_{n}),\label{env3}\\
\eta^{*}(x,n;\{z^{*}\}^{n}_{n})&=& {\sum_{k}\/}'\eta^{*}_{k}(n)
\psi^{*}_{k}(x;\{z^{*}\}^{n}_{n}),\label{env4}
\eeqa
\seeqa
where $\sum'_{k}$ denotes 
summation over the modes excluding the zero-modes. 
The zero-modes function is proportional to 
the first derivative of $\xi^{{\rm s}}(x;z(n))$ 
with respect to $x$;
\beqa
\psi_{0}(x;z(n)) = -A\lambda\frac{d\xi^{{\rm s}}(x;z(n))}{dx},
\label{zero-mode}
\eeqa
where $A$ is a real normalization constant. 
To confirm this relation, one may consider 
the sum of the domain wall configuration 
and the zero-mode part:
\beqa
&&\xis(x;z(n))+\eta_{0}(n)\psi_{0}(x;z(n))\non\\
&&=\xis(x;z(n))-A\lambda\eta_{0}(x)\frac{d\xis(x;z(n))}{dx}\non\\
&&\simeq \xis(x-A\lambda\eta_{0}(n);z(n))\non\\
&&=\exp\left[-
\frac{x}{\lambda}+q(n)+A\Re\eta_{0}(n)
+i(\phi_{0}(n)+A\Im{\eta}_{0}(n))\right].
\eeqa
Hence, it is clear that Eq.(\ref{zero-mode}) 
gives the zero-mode eigen function, 
and that the real and the imaginary parts of 
$\eta_{0}(n)$ are the zero-modes 
corresponding to the translation and the rotation modes, respectively. 
We choose $\{\psi_{k}\}$ so that 
$\{\psi_{0},\psi_{k}\}$ 
forms a orthonormal set;
\sbeqa
\label{n-o}
\beqa
&&\int_{-L/2}^{L/2}\frac{dx}{a}f(x;\{z\}^{n}_{n})\psi^{*}_{0}(x;z^{*}(n))
\psi_{0}(x;z(n))=1,\label{n-o1}\\
&&\int_{-L/2}^{L/2}\frac{dx}{a}f(x;\{z\}^{n}_{n})\psi^{*}_{0}(x;z^{*}(n))
\psi_{k}(x;\{z\}^{n}_{n})=0,\label{n-o2}\\
&&\int_{-L/2}^{L/2}\frac{dx}{a}f(x;\{z\}^{n}_{n})\psi^{*}_{k}
(x;\{z^{*}\}^{n}_{n})
\psi_{k'}(x;\{z\}^{n}_{n})=\delta_{kk'},\label{n-o3}\\
&&f(x;\{z\}^{n}_{n})\bigg\{\psi_{0}(x;z(n))\psi^{*}_{0}(x;z^{*}(n)) \non\\
&&+ {\sum_{k}\/}'\psi_{k}(x;\{z\}^{n}_{n})\psi^{*}_{k}(x';\{z^{*}\}^{n}_{n})
\bigg\} = \delta\left(\frac{x-x'}{a}\right)\label{n-o4}.
\eeqa
\seeqa
where $f(x;\{z\}^{n}_{n})$ is a real weight function to be 
fixed later. 
This weight function neglects a non-linear character 
of the coherent state path integral.

\subsection{Faddeev-Popov type identity}
In order to introduce the collective degrees of freedom and eliminate
the zero modes, we adopt the Faddeev-Popov method  
to the spin-coherent state path integral. 
To do so, we invoke the following Faddeev-Popov type identity:
\sbeqa
\beqa
1&=&\int
dRg_{0}^{[\xi,\xi^{*}]}dIg_{0}^{[\xi,\xi^{*}]}
\delta(Rg_{0}^{[\xi,\xi^{*}]})
\delta(Ig_{0}^{[\xi,\xi^{*}]})\non\\
&=&
\int\frac{dz(n)dz^{*}(n)}{2i}
\Delta^{[\xi,\xi^{*}]}(\{z\}^{n}_{n})
\delta(Rg_{0}^{[\xi,\xi^{*}]}(\{z\}^{n}_{n}))
\delta(Ig_{0}^{[\xi,\xi^{*}]}(\{z\}^{n}_{n})),
\label{F-P}
\eeqa
where 
\beqa
\frac{dz(n)dz^{*}(n)}{2i}=dq(n)d\phi_{0}(n), 
\eeqa
\seeqa
and $\Delta^{[\xi,\xi^{*}]} (\{z\}^{n}_{n})$ is the Faddeev-Popov type
determinant 
\beqa
\Delta^{[\xi,\xi^{*}]} (\{z\}^{n}_{n})
=\left|
\frac{\partial(g_{0}^{[\xi]}(\{z\}^{n}_{n}),g_{0}^{*[\xi^{*}]}
(\{z^{*}\}^{n}_{n}))}
{\partial(z(n),z^{*}(n))}
\right|,
\label{F-P-D2}
\eeqa
with $Rg_{0}^{[\xi,\xi^{*}]}(\{z\}^{n}_{n})$ and
$Ig_{0}^{[\xi,\xi^{*}]}(\{z\}^{n}_{n})$
being the real and the imaginary parts of  
$g_{0}^{[\xi]}(\{z\}^{n}_{n})$ defined as 
\sbeqa 
\label{F-P-g}
\beqa
g_{0}^{[\xi]}(\{z\}^{n}_{n})
&:=& \int_{-L/2}^{L/2}\frac{dx}{a}f(x;\{z\}^{n}_{n})
\{\xi(x,n)-\xi^{{\rm s}}(x;z(n))\}\psi^{*}_{0}(x;z^{*}(n)),
\label{F-P-g1}\\
g_{0}^{*[\xi^{*}]}(\{z^{*}\}^{n}_{n})
&:=& \{g_{0}^{[\xi]}(\{z\}^{n}_{n}) \}^{*}.
\label{F-P-g2}
\eeqa
\seeqa
\subsection{Transition amplitude}

Hereafter we consider transition amplitudes between domain-wall states. 
Namely we take 
\beqa
|\xi_{\beta}\>=|z_{\beta}\>
:=\bigotimes_{j}^{N_{L}}|\xis(ja;z_{\beta})\>,\qquad \beta=I,F.
\eeqa
in Eq. (\ref{ta1}). 
Inserting the identity (\ref{F-P}) 
into the right-hand side of (\ref{ta1}) at each discrete time, we get
\beqa
&&\<z_{F}|e^{-i\hat{H}T/\hbar}|z_{I}\>\non\\
&&=
\lim_{N\to\infty}\prod_{n=1}^{N-1}\int\prod_{x}
\frac{d\xi(x,n)d\xi^{*}(x,n)}{2\pi i}\frac{2S+1}
{(1+\xi^{*}(x,n)\xi(x,n))^{2}}\non\\
&&\times\left\{
\int\prod_{n=1}^{N-1}\frac{dz(n)dz^{*}(n)}{2i}
\Delta^{[\xi,\xi^{*}]}(\{z\}^{n}_{n})
\delta(Rg_{0}^{[\xi,\xi^{*}]}(\{z\}^{n}_{n}))
\delta(Ig_{0}^{[\xi,\xi^{*}]}(\{z\}^{n}_{n}))
\right\}\non\\
&&\times\exp\left(
\frac{i}{\hbar}{\cal S}[\xi^{*},\xi]
\right).
\label{F-P-ta1}
\eeqa
At this stage, we substitute (\ref{xi-c-e}) with (\ref{n-o}) for 
$\xi$ and $\xi^{*}$. 
Then (\ref{F-P-g}) may be replaced by 
\beqa
g_{0}^{[\xis+\tilde{\eta}]}(\{z\}^{n}_{n})= \eta_{0}(n),\qquad 
g_{0}^{*[\bxis+\tilde{\eta}^{*}]}(\{z^{*}\}^{n}_{n})= \eta^{*}_{0}(n).
\eeqa
Thus,
\beqa
Rg_{0}^{[\xis+\tilde{\eta},\bxis+\tilde{\eta}^{*}]}(\{z\}^{n}_{n})=
\Re \eta_{0}(n),\qquad 
Ig_{0}^{[\xis+\tilde{\eta},\bxis+\tilde{\eta}^{*}]}(\{z\}^{n}_{n})=
\Im \eta_{0}(n),
\eeqa
which are zero-mode degrees of freedom. 
The integration measure can be rewritten as 
\beqa
&&\prod_{x}
\left.
\frac{d\xi(x,n)d\xi^{*}(x,n)}{2\pi 
i}\frac{2S+1}{(1+\xi^{*}(x,n)\xi(x,n))^{2}}
\right|_{\xi=\xis+\tilde{\eta},\xi^{*}=\bxis+\tilde{\eta}^{*}}\non\\
&=&
\prod_{k=0}\frac{d\eta_{k}(n)d\eta^{*}_{k}(n)}{2\pi i}2S\left(1+\frac{1}{2S}\right)
\prod_{x}
\Xi[x,n;z,\tilde{\eta}]
,\label{int-measure}
\eeqa
where
\beqa 
\Xi[x,n;z,\tilde{\eta}] 
:=
\frac{\{ f(x;\{z\}^{n}_{n})\}^{-1} }{\{           
1+(\bxis(x;z^{*}(n))+\tilde{\eta}^{*}(x;\{z^{*}\}^{n}_{n}))(\xis(x;z(n))+\tilde{\eta}(x;\{z\}^{n}_{n}) )
\}^{2} }.
\label{Xi}
\eeqa
Thus, the transition amplitude (\ref{F-P-ta1}) reduced to 
\beqa
\<z_{F}|e^{-i\hat{H}T/\hbar}|z_{I}\>
&=&
\lim_{N\to\infty}\int\prod_{n=1}^{N-1}
\Bigg[
\frac{dz(n)dz^{*}(n)}{2i}
\prod_{k=0}\frac{d\eta_{k}(n)d\eta^{*}_{k}(n)}{2\pi i}
2S\left(1+\frac{1}{2S}\right)\non\\
&&\times\prod_{x}\Xi[x,n;z,\tilde{\eta}]\cdot
\Delta^{[\xis+\tilde{\eta},\bxis+\tilde{\eta}^{*}]}
(\{z\}^{n}_{n})\delta(\Re \eta_{0}(n))\delta(\Im \eta_{0}(n))\non\\
&&\times
\exp\left(
\frac{i}{\hbar}{\cal S}[\bxis+\tilde{\eta}^{*},\xis+\tilde{\eta}]
\right)\Bigg].
\label{F-P-ta2}
\eeqa
With the help of the delta function, 
we can immediately integrate out the zero-modes 
at each discrete time. 
Furthermore, in the case of large spin $S \gg 1$, 
The factors contributing to the integration measure takes 
the simple form [19]:
\sbeqa
\beqa
&&\Xi[x,n;z,\eta]
=1+\order(S^{-1/2}),\\
&&\Delta^{[\xis+{\eta},\bxis+{\eta}^{*}]}(\{z\}^{n}_{n})
=\frac{N_{{\rm DW}}}{2}\left(1+\order(S^{-1/2})\right).
\eeqa
\seeqa
In this way we find 
\beqa
\<z_{F}|e^{-i\hat{H}T/\hbar}|z_{I}\>
&=&
\lim_{N\to\infty}\int\prod_{n=1}^{N-1}
\Bigg[
{\prod_{k}\/}'2S\frac{d\eta_{k}(n)d\eta^{*}_{k}(n)}{2\pi i}
\cdot N_{{\rm DW}}S\frac{dz(n)dz^{*}(n)}{2\pi i}\non\\
&&\times\exp\left(
\frac{i}{\hbar}{\cal S}[\bxis+{\eta}^{*},\xis+{\eta}]
\right)\Bigg].
\label{F-P-ta3}
\eeqa
This formula determines the quantum dynamics of the domain wall 
in the path integral.

\subsection{Effective Action}

The next step is to expand the action 
${\cal S}[\bxis+{\eta}^{*},\xis+{\eta}]$ with respect to the environment. 
The expanded action consists of three parts.
The first part ${\cal S}_{{\rm c}}[z^{*},z]$ contains only 
the collective degrees of freedom. 
This part was estimated in detail in Ref.[16]. 
The second part ${\cal S}_{{\rm e}}[\eta^{*},\eta]$ contains only 
the environment, and the third part 
${\cal S}_{{\rm c-e}}[z^{*},z,\eta^{*},\eta]$ 
represents the interaction between the collective 
degrees of freedom and the environment; 
\beqa
{\cal S}[\bxis+{\eta}^{*},\xis+{\eta}] 
:= {\cal S}_{{\rm c}}[z^{*},z]
+ {\cal S}_{{\rm e}}[\eta^{*},\eta]
+ {\cal S}_{{\rm c-e}}[z^{*},z,\eta^{*},\eta].
\eeqa

Now, we proceed to find on equation to determine the environmental mode 
functions $\{\psi_{k}\}$. 
Such an equation may be obtained from the dynamical term 
whose collective part gave 
the static equation(\ref{stat-eq}). 
Accordingly, 
we examine the dynamical term of ${\cal S}_{{\rm e}}[\eta^{*},\eta]$, 
which we denote by ${\cal S}^{{\rm d}}_{{\rm e}}[\eta^{*},\eta]$: 
\sbeqa
\beqa
{\cal S}^{{\rm d}}_{{\rm e}}[\eta^{*},\eta] &:=& 
-2KS\left(S-\frac{1}{2}\right)\sum_{n=1}^{N}\epsilon
\int_{-L/2}^{L/2}\frac{dx}{a}\eta^{*}(x,n;\{z^{*}\}^{n}_{n})
g^{2}(x;\{z\}^{n}_{n})\non\\
&&\times
\Bigg[-\lambda^{2}\frac{\partial^{2}}{\partial x^{2}}
-4\lambda g(x;\{z\}^{n}_{n})
\exp(-2\tilde{x}^{n}_{n-1})\frac{\partial}{\partial x}\non\\
&&-\bigg\{3\exp(-2\tilde{x}^{n}_{n-1})-1\bigg\}g(x;\{z\}^{n}_{n})\Bigg]
\eta(x,n-1;\{z\}^{n}_{n-1}),
\label{ex-action-ed1}
\eeqa
where 
\beqa
\tilde{x}^{n}_{m} =\tilde{x}^{*m}_{n}\equiv 
\frac{x}{\lambda}-\frac{z^{*}(n)+z(m)}{2},
\qquad
g(x;\{z\}^{n}_{n})\equiv
\frac{1}{1+e^{-2\tilde{x}^{n}_{n-1}}}.
\eeqa
\seeqa
Terms with proportional to $\eta^{*}\eta^{*}$ or $\eta\eta$ vanish 
due to the static equation(\ref{stat-eq}). 
Note that 
this cancellation is not a generally guaranteed theorem. 
It turns out to be conevenient to write 
$\{\psi_{k},\psi^{*}_{k}\}$ in (\ref{env3}) and (\ref{env4}) 
as follows:
\sbeqa
\label{eigen-function}
\beqa
\psi_{k}(x;\{z\}^{n}_{n})&:=&
2e^{-x/\lambda + z(n)}\cosh\left(
\tilde{x}^{n}_{n}
\right)
\varphi_{k}\left(\tilde{x}^{n}_{n}\right),
\label{eigen-function1}\\
\psi^{*}_{k}(x;\{z^{*}\}^{n}_{n})&:=& \{\psi_{k}(x;\{z\}^{n}_{n}) \}^{*}.
\label{eigen-function2}
\eeqa
\seeqa
Putting (\ref{env3}), (\ref{env4}) and (\ref{eigen-function1}) 
into (\ref{eigen-function2}), we find 
\beqa
&&{\cal S}^{d}_{e}[\eta^{*},\eta]
=-2\lambda^{2}KS\left(S-\frac{1}{2}\right)
\sum_{n=1}^{N}\epsilon
\sum_{k',k}\eta^{*}_{k'}(n)\eta_{k}(n-1)\non\\
&\times&
\int_{-L/2}^{L/2}\frac{dx}{a}
\varphi^{*}_{k'}(x)
\Bigg[
-\frac{\partial^{2}}{\partial x^{2}}+
\frac{1}{\lambda^{2}}
\bigg\{1-2{\rm sech}^{2}\left(\frac{x}{\lambda}\right)
\bigg\}\Bigg]\varphi_{k}(x)
\eeqa
where we replaces $x/\lambda-(z^{*}(n)+z(n))/2$ by $x/\lambda$ 
because $L/\lambda \gg  1$. 
Thus, we are led to demand $\varphi_{k}(x)$ 
to obey the eigenvalue equation  
\beqa
\Bigg[
-\partial_{x}^{2}+
\frac{1}{\lambda^{2}}
\bigg\{1-2{\rm sech}^{2}
\left(\frac{x}{\lambda}\right)\bigg\}
\Bigg]\varphi_{k}(x)
=\omega_{k}\varphi_{k}(x),
\eeqa
where $\omega_{k}$ is the eigenvalue. 
This solution to this equation 
is well known in the literature; 
\sbeqa
\beqa
\varphi_{k}(x)=
N_{k}\left(-ik\lambda+\tanh(x/\lambda)\right) e^{ikx},
\eeqa
where 
\beqa
N_{k}=\left(\frac{L}{a}(k^{2}\lambda^{2}+1)\right)^{-1/2},\qquad
\omega_{k}=k^{2}+\lambda^{-2}.
\eeqa
\seeqa
These eigenfunctions giagonalize 
${\cal S}^{{\rm d}}_{{\rm e}}[\eta^{*},\eta]$ as 
\sbeqa
\beqa
{\cal S}^{d}_{e}[\eta^{*},\eta]
=-2S\sum_{n=1}^{N}{\sum_{k}\/}'\hbar
\epsilon\Omega_{k}\eta^{*}_{k}(n)\eta_{k}(n-1),
\label{ex-action-ed3}
\eeqa
where
\beqa
\Omega_{k}\equiv \frac{JS}{\hbar}(k^{2}+\lambda^{-2}).
\eeqa
\seeqa
The last equation 
is the dispersion relation of 
the environmental eigenmode. 
Note that $\omega_{k}$ is not 
zero in the limit of $k\to0$.

What remains is to decide the weight function 
$f(x;\{z\}^{n}_{n})$.  
With (\ref{eigen-function}), the condition (\ref{n-o}) takes 
the form 
\beqa
\int_{-L/2}^{L/2}\frac{dx}{a}
f(x,\{z\}^{n}_{n})4
e^{-2\tilde{x}^{n}_{n}}\cosh^{2}
\left(
\tilde{x}^{n}_{n}
\right)
\varphi^{*}_{k}(x;\{z^{*}\}^{n}_{n})\varphi_{k'}(x;\{z\}^{n}_{n})=\delta_{k,k'}.
\eeqa
This is satisfied if we choose 
\beqa
f(x,\{z\}^{n}_{n})=
\frac{1}{(1+
e^{-2x/\lambda + z^{*}(n)+z(n)})^{2}}.
\label{weight-f}
\eeqa
It is easy to check that (\ref{n-o2}) is also satisfied by this 
choice. 
Accordingly, 
the normalization constant $A$ of 
the zero-mode function $\psi_{0}(x;z(n))$ as 
given by (\ref{zero-mode}) may be determined by use of 
(\ref{weight-f}) and (\ref{n-o1}):
\beqa
1=A^{2}
\int_{-L/2}^{L/2}\frac{dx}{a}\frac{e^{-2x/\lambda +
z^{*}(n)+z(n)}}{(1+e^{-2x/\lambda + z^{*}(n)+z(n)})^{2}}\non
&=&\frac{A^{2}\lambda}{4a}
\left[
\tanh
\left(\tilde{x}^{n}_{n}\right)
\right]_{-L/2\to -\infty}^{L/2\to\infty}\non\\
&=&\frac{A^{2}\lambda}{2a}.
\eeqa
Thus,
\beqa
A=
\sqrt{\frac{2a}{\lambda}}.
\eeqa
The closure (\ref{n-o4}) can also be shown to hold with these choices. 

The remainder of the action can 
be calculated 
in the same manner [19], and we obtain the effective action 
in the following form: 
\sbeqa
\beqa
&&\frac{i}{\hbar}{\cal S}_{{\rm s}}[z^{*},z]=
N_{{\rm DW}}S\sum_{n=1}^{N}
\bigg[
-\frac{1}{2}\bigg\{z^{*}(n)z(n)+z^{*}(n-1)z(n-1)\bigg\}\non\\
&&+z^{*}(n)z(n-1)
-\frac{i}{\hbar}E_{{\rm DW}}T\bigg],
\label{action-s}\\
&&\non\\
&&\frac{i}{\hbar}{\cal S}_{{\rm e}}[\eta^{*},\eta]=
2S\sum_{n=1}^{N}{\sum_{k}\/}'
\bigg[
-\frac{1}{2}\bigg\{
\eta^{*}(n)\eta(n)+\eta^{*}(n-1)\eta(n-1)\bigg\}\non\\
&&+\eta^{*}(n)\eta(n-1)
-\frac{i}{\hbar}\epsilon\Omega_{k}\eta^{*}(n)\eta(n-1)
\bigg],
\label{action-env}\\
&&\non\\
&&\frac{i}{\hbar}{\cal S}_{{\rm c-e}}[z^{*},z,\eta^{*},\eta]=
2S\sum_{n=1}^{N}{\sum_{k}\/}'\bigg[
\bigg(J_{k}(\Delta z(n))\eta^{*}_{k}(n)-J_{k}(\Delta z^{*}(n+1))\eta_{k}(n)\bigg)\non\\
&&+ \order(S^{-1})\bigg],
\label{action-s-env}\\
&&\non\\
\eeqa
where $\Delta z(n) \equiv z(n)-z(n-1)$, and 
\beqa
J_{k}(z) := \frac{\lambda}{2a}\int_{-L/2\lambda}^{L/2\lambda}
dx\frac{\sinh\frac{z}{2}}{\cosh\left(x+\frac{z}{2}\right)}\varphi_{k}^{*}(x),
\eeqa
\seeqa
which is a function of the collective 
degrees of freedom, represents a non-linear interaction with 
the environment. 
It is seen that the above form of the effective action is 
formally the same as that obtained in a 
boson coherent state path
integral. Hence, it is concluded that 
the quantum dynamics of the domain wall is formally represented by 
that of a "boson" $z$ interacting with environmental bosons $\{\eta\}$. 
However, it is to be remenbered that 
the imaginary part of the "boson" $z$ is an 
angular variable. 
This circumstance can cause a subtle departure from the case of a boson. 
Details of these subtleties as well as 
a concrete evaluation of transition amplitudes are left for a future work.

\newcommand{\etal}{{\em et al.}}
\setlength{\parindent}{0mm}
\vspace{5mm}
{\bf References}
\begin{list}{}{\setlength{\topsep}{0mm}\setlength{\itemsep}{0mm}%
\setlength{\parsep}{0mm}}
%
%
\item[1.] P. C. E. Stamp, E. M. Chudnovsky, and B. Barbra, Int. J. Mod. Phys. B {\bf 6}, 1355 (1992).

\item[2.] {\it Quantum Tunneling of Magnetization}, Proceedings of the NATO workshop, Chichilianne, France, 1994, edited by L. Gunther and B. Barbara (Kluwer Academic, Norwell, MA, 1995).

\item[3.] E. M. Chudnovsky and J. Tejada, {\it Macroscopic Quantum Tunneling of the Magnetic Moment} (Cambridge University Press, United Kingdom, 1998).

\item[4.] T. Egami, Phys. Status Solidi B {\bf 57}, 211 (1973); Phys. Status Solidi A {\bf 19}, 747 (1973); {\bf 20}, 157 (1973).

\item[5.] P. C. E. Stamp, Phys. Rev. Lett. {\bf 66}, 2802 (1991).

\item[6.] E. M. Chudnovsky, O. Iglesias, and P. C. E. Stamp, Phys. Rev. B {\bf 46}, 5392 (1992).

\item[7.] G. Tatara and H. Fukuyama, Phys. Rev. Lett. {\bf 72}, 772 (1994); J. Phys. Soc. Jpn. {\bf 63}, 2538 (1994).

\item[8.] H. B. Braun and D. Loss, Phys. Rev. B {\bf 53}, 3237 (1996).

\item[9.] H. B. Braun and D. Loss, J. Appl. Phys. {\bf 79}, 6107 (1996).

\item[10.] S. Takagi and G. Tatara, Phys. Rev. B {\bf 54}, 9920 (1996).

\item[11.] B. A. Ivanov, A. K. Kolezhuk, and V. E. Kireev, Phys. Rev. B {\bf 58}, 11514 (1998).


\item[12.] J. R. Klauder, Phys. Rev. D {\bf 19}, 2349 (1979).

\item[13.] H. G. Solari, J. Math. Phys. {\bf 28}, 1097 (1987).

\item[14.] E. A. Kochetov, J. Math. Phys. {\bf 36}, 4667 (1995).

\item[15.] J. Shibata and S. Takagi, Int. J. Mod. Phys. B {\bf 13}, 107 (1999).

\item[16.] J. Shibata and S. Takagi, Phys. Rev. B {\bf 62}, 5719 (2000). 

\item[17.] J. M. Radcliffe, {\em J. Phys. } {\bf A 4}, 313 (1971).

\item[18.] R. Rajaraman, {\it Solitons and Instantons} (North-Holland, Amsterdam, 1982).

a\item[19.] J. Shibata, Doctor Thesis in Tohoku Unversity, 2001.

\end{list}

\end{document}